\documentclass{article}
\usepackage{amsmath}

\title{Nonperturbative operator quantization of strongly 
nonlinear fields}
\author{V. Dzhunushaliev
\thanks{E-mail: dzhun@hotmail.kg}}

\date{}

\begin{document}

\maketitle

\begin{center}
\textit{
Dept. Phys. and Microel. Engineer., Kyrgyz-Russian
Slavic University, Bishkek, Kievskaya Str. 44, 720000, Kyrgyz
Republic}
\end{center}

\begin{abstract}
At present an algebra of strongly interacting fields 
is unknown. In this paper it is 
assumed that the operators of strongly nonlinear field can form 
a non-associative algebra. It is shown that such algebra can be described 
as an algebra of some pairs. The comparison of presented techniques 
with the Green's functions method in the superconductivity theory is made. 
A possible application to the QCD and High-T$_c$ superconductivity theory 
is discussed.
\end{abstract}

\section{INTRODUCTION}

The quantization rules in quantum field theory are applied for 
the noninteracting part of Lagrangian only and then the n-point Green's
functions are calculated using of the Feynman diagram technique.
Such procedure is not valid for theories with strong nonlinear
fields, for example, for the quantum chromodynamics (QCD) and gravity.
This means that in QCD we have a hypothesized flux tube stretched between
quark and antiquark and such nonlocal object can not be explained 
by the use of the perturbative diagram techniques. In the 50's 
Heisenberg conceived the difficulties of using an expansion in
small parameter for quantum field theories with strong interactions 
(see, for example, Ref's \cite{heis1,heis2}). In these papers it was 
repeatedly underscored that a nonlinear theory with a strong
coupling requires the introduction of another quantization
procedure. Heisenberg's basic idea proceeds from the fact that the
n-point Green's functions must be found from some infinite set of
differential equations derived from the field equations for the
field operators. In this case the n-point Green's functions and
the propagator are not connected with one another by a simple manner 
as it is in the case of diagram techniques. Later this idea has been 
abandoned because of big mathematical problems connected to an obtained 
infinite equations set. Nevertheless it can be shown \cite{dzh3} that 
the Green's function method and Ginzburg - Landau equation 
in superconductivity theory is actually a realization of the 
Heisenberg's quantization procedure.
\par
For the standard Feynman diagram techniques it is very important
that the corresponding classical field equations have wave solutions which
become quanta after quantization. But for the strongly nonlinear
field theory the situation can be drastically changed : such theory can 
have (on the classical level) such (possible singular) solutions 
which after quantization become some nonlocal quantum objects 
(for example, it can be the flux tube or monopole-like 
configuration \cite{dzh2}). In this case a nonlinearity leads to 
a non-locality, \textit{i.e.} not every quantum
object can be described as a cloud of quanta.
\par
In this situation (for strongly nonlinear fields) we have a
question : what is the algebra of quantized fields~? We know the
answer for the noninteracting fields only : it is a noncommutative
algebra with canonical commutation relations. The assumption
presented here is : in the case of strongly nonlinear fields we should
change the algebra of quantized fields to derive 
functions, which can be connected with the n-point Green's
functions. In this paper we suppose that it can be done if the
algebra of quantized fields be much more complicated then 
for noninteracting fields, for example, it can be 
a non-associative algebra \cite{wulk}.
\par 
Ordinarily the non-associative algebra was used in classical 
field theories \cite{Kurd:1985bj}, \cite{Lohmus:1984ny}, in quantum 
gravity \cite{Nesterov:2000qb} to describe a discrete 
spacetime and in a non-associative geometry. In contrast to these 
approaches the idea, presented here, consists of following regulations 
\begin{enumerate}
  \item 
  The non-associative algebra (of strongly nonlinear quantum fields) can 
  give additional degrees of freedom that is relevant to the description 
  of the n-point Green's functions which can not be calculated on the language 
  of Feynman diagrams.
  \item 
  The Heisenberg's quantization procedure \cite{heis1,heis2} gives us 
  information to define of the algebra of strongly nonlinear 
  quantum fields.
\end{enumerate}

\section{ALGEBRA OF QUANTIZED FIELDS \\ AS A NON-ASSOCIATIVE ALGEBRA}

Let us assume that an algebra of quantized fields is a
non-associative algebra \cite{dzh1}. It means that
\begin{equation}
\label{sec1-1}
  a \Bigl( bc \Bigl) \neq \Bigl( ab \Bigl) c
\end{equation}
here the operators $a,b,c$ are the operators of quantized fields.
Let us introduce an associator $Ass(a,b,c)$
\begin{equation}
\label{sec1-2}
  \Bigl( ab \Bigl)c - a\Bigl( bc \Bigl) = Ass(a,b,c).
\end{equation}
In the general case $Ass(a,b,c)$ can be an operator. For example, we can 
calculate following commutator
\begin{equation}
\begin{split}
\label{sec1-3}
  \Bigl(ab\Bigl) \Bigl(cd\Bigl) - \Bigl(cd\Bigl) \Bigl(ab\Bigl) = 
  \Bigl(c\Delta_{ad} \Bigl)b +
  \Delta_{ac}\Bigl(db\Bigl) + a\Bigl(c\Delta_{bd} \Bigl) +
  a\Bigl(\Delta_{bc}d \Bigl) \\
  + \Bigl(Ass(c,a,d) - Ass(c,d,a)\Bigl)b +
  a\Bigl(Ass(c,b,d) - Ass(b,c,d)\Bigl) \\
  + Ass(cd,a,b) - Ass(ca,d,b) + Ass(a,b,cd) - Ass(a,c,db)
\end{split}
\end{equation}
here $\Delta_{ab} = ab - ba$ is a commutator that can be an operator. 
More detailed derivation of this equation see in Appendix \ref{app1}. 
\par
The non-associative algebra can be an alternative algebra,
\textit{i.e.}
\begin{equation}
\label{sec1-4}
  a(ab) = (a^2)b = a^2 b \qquad \text{and/or} \qquad
  (ba)a = b(a^2) = b a^2 .
\end{equation}
With $b = a = \varphi (x)$ and $c = d = \varphi^*(y)$, where 
$\varphi^*(y)$ is a Hermitian-conjugated operator. 
Eq. \eqref{sec1-3} has following form
\begin{equation}
\begin{split}
\label{sec1-5}
  \Bigl (\varphi(x)\varphi(x) \Bigl )
  \Bigl (\varphi^*(y)\varphi^*(y) \Bigl ) -
  \Bigl (\varphi^*(y)\varphi^*(y) \Bigl )
  \Bigl (\varphi(x)\varphi(x) \Bigl ) \\
  = 2 \Bigl (\varphi(x) \varphi^*(y) +
  \varphi^*(y) \varphi(x)\Bigl ) \Delta (x,y) -
  2\varphi^*(y) 
  Ass\left(\varphi(x),\varphi^*(y),\varphi(x)\right) \\ 
  + Ass\left(\varphi^*(y),\varphi^2(x),\varphi^*(y)\right) . 
  \end{split}
\end{equation}
Here we suppose that the commutator 
$\Delta(x,y) = \varphi(x)\varphi^*(y) - \varphi^*(y)\varphi(x)$, 
associators 
$Ass\left(\varphi(x),\varphi^*(y),\varphi(x)\right) = 
\Bigl(\varphi(x)\varphi^*(y)\Bigl)\varphi(x) -
\varphi(x)\Bigl(\varphi^*(y)\varphi(x)\Bigl)$ and
$Ass\left(\varphi^*(y),\varphi^2(x),\varphi^*(y)\right) = 
\Bigl(\varphi^*(y)\varphi^2(x)\Bigl) \varphi^*(y) -
\varphi^*(y)\Bigl(\varphi^2(x) \varphi^*(y)\Bigl)$
are $c-$numbers. The first terms of the right-hand side 
of Eq.\eqref{sec1-3} 
is similar to the terms, appearing by time ordering of the canonical 
quantization of noninteracting fields.
\par
A similar construction can be written for fermions
\begin{equation}
\begin{split}
\label{sec1-6}
  \Bigl (\psi_\alpha(x)\psi_\alpha(x) \Bigl )
  \Bigl (\psi^+_\beta(y)\psi^+_\beta(y) \Bigl ) -
  \Bigl (\psi^+_\beta(y)\psi^+_\beta(y) \Bigl )
  \Bigl (\psi_\alpha(x)\psi_\alpha(x) \Bigl ) \\
  = -2 \Delta^2 \left(\psi_\alpha(x), \psi^+_\beta(y) \right) + 
  2 \psi_\alpha(x) Ass\left(\psi^+_\beta(y)\psi_\alpha(x), 
  \psi^+_\beta(y), \psi_\alpha(x) \right) \\
  - Ass\left(\psi_\alpha(x), \psi^+_\beta(y), \psi^+_\beta(y)
  \psi_\alpha(x) \right)
\end{split}
\end{equation}
here $\Delta \left(\psi_\alpha(x), \psi^+_\beta(y) \right) = 
\psi_\alpha(x)\psi^+_\beta(y) + \psi^+_\beta(y) \psi_\alpha(x)$ 
is the anticommutator. 
The most interesting things in Eq's.\eqref{sec1-5} \eqref{sec1-6} are the 
last terms $Ass(\varphi^*(y),\varphi^2(x),\varphi^*(y))$ and 
$Ass\left(\psi_\alpha(x), \psi^+_\beta(y), \psi^+_\beta(y)
\psi_\alpha(x) \right)$. 
On the left-hand side of these equations we
have commutators. As well as in the superconductivity theory, 
the operator $\psi^+_\beta(y)\psi^+_\beta(y)$ can be interpreted as 
a creation of a pair in the point $y$ (it can be Cooper 
pair in the superconductivity theory or quark-antiquark pair 
in the QCD while the distance between fermions tends to zero) 
and consequently the operator $\psi_\alpha(x)\psi_\alpha(x)$ 
describes an annihilation of the pair in the point $x$. 
The first term of the right-hand side of Eq. 
\eqref{sec1-5} \eqref{sec1-6} is the ordinary term appearing
in noninteracting fields. If all quantum particles are
combined into pairs (for example, it happens in the ground state
of a superconductor for $T = 0$ when all electrons are coupled into 
Cooper pairs) then $\langle \varphi^*(x)\rangle = 0$ 
and $\langle \psi_\alpha(x) \rangle = 0$ and second terms 
of right-hand side of Eqs. \eqref{sec1-5} \eqref{sec1-6} are 
equal to zero after quantum averaging. 
In this case the last terms in these equations have 
clear meaning : the origin of this term is the nonlinearity
of the field that describes a propagation of a pair like to 
Cooper pair in the superconductivity or a quark - antiquark 
pair in the QCD. Let us compare these equations 
with the expression of average of four $\psi-$~operators 
in the Green's function method in the superconductivity
theory \cite{abr} 
\begin{equation}
\begin{split}
    \left \langle T \left ( \hat\psi_\alpha (x_1) \hat\psi_\beta (x_2)
    \hat\psi^+_\gamma (x_3) \hat\psi^+_\delta (x_4)
                \right )
    \right \rangle \\ 
    \approx -\left \langle T \left ( \hat\psi_\alpha (x_1)
    \hat\psi^+_\gamma (x_3)
                 \right )
    \right \rangle \left \langle T \left ( \hat\psi_\beta (x_2) \hat\psi^+_\delta
    (x_4)
                 \right )
    \right \rangle  \\ 
    + \left \langle T \left ( \hat\psi_\alpha (x_1)
    \hat\psi^+_\delta (x_4)
                 \right )
    \right \rangle \left \langle T \left ( \hat\psi_\beta (x_2) \hat\psi^+_\gamma
    (x_3)
                 \right )
    \right \rangle  \\ 
    + \left \langle N \left\vert T \left ( \hat\psi_\alpha (x_1)
    \hat\psi_\beta (x_2)
    \right ) \right\vert N + 2
    \right \rangle \left \langle N +2 \left\vert T \left ( \hat\psi^+_\gamma (x_3)
    \hat\psi^+_\delta (x_4)
    \right ) \right\vert N
    \right \rangle
\end{split}
\label{sec1-7}
\end{equation}
where $|N \rangle$ and $|N + 2 \rangle$ are ground states of system with $N$
and $N +2$ particles (Cooper pairs), respectively.
Comparing this expression with Eq.\eqref{sec1-6}, we see immediately that the
last term in Eqs. \eqref{sec1-5} \eqref{sec1-6} can be interpreted as a
propagator for the pairs $(\varphi(x)\varphi(x))$ or 
$(\psi_\alpha(x)\psi_\alpha(x))$, \textit{i.e.} this 
propagator is connected with the associator. In other words in some situation 
the nonlinear interaction leads to the appearance of pairs 
and the non-associative algebra of quantum fields 
can describe the propagation of these pairs. 
In Ref. \cite{dzh3} it is shown that the above-mentioned 
Green's function method in the superconductivity theory realizes 
the Heisenberg's idea about the calculation of the Green's 
functions for fields with a nonlinear interaction. 
In this case the non-associative algebra can give us a 
possibility to describe field operators with 
the nonlinear
interaction.

\section{``COOPER PAIRING'' \\ IN NON-ASSOCIATIVE ALGEBRA}

Let show that algebra of quantized fields mentioned above 
can have interesting properties similar to the formation of
fermion pairs in the superconductivity theory and in the QCD.
In the first case it is a Cooper pair containing two electrons
are connected with one another by phonons and in the second case 
it is the quark and antiquark held by 
a hypothesized flux tube filled with the SU(3) gauge field.
\par
Let us assume that the commutator/anticommutator $\Delta_{ab}$ and
associator $Ass(a,b,c)$ are $c-$numbers for any operators $a,b,c$.
This means that
\begin{eqnarray}
    \varphi (x) \varphi^* (y)  \mp \varphi^* (y) \varphi (x)
    & = & \Delta (x,y) ,
    \label{sec2-1}\\
    \Bigl(a(x) b(y) \Bigl)c(z) -
    a(x) \Bigl (b(y) c(z)\Bigl)
    & = & Ass\left(a(x),b(y),c(z)\right)
\label{sec2-2}
\end{eqnarray}
here the operators $a,b,c$ can be changed by the operators
$\varphi (x), \varphi (y), \varphi (z)$ or 
$\varphi^* (x), \varphi^* (y), \varphi^* (z)$; the operator 
$\varphi(x) $ can have some indices but it is inessential 
for us; the sign $(+)$ in Eq. \eqref{sec2-1} is connected with a spinor field. 
Thus, $\Delta (x,y)$ and $Ass(a(x),b(y),c(z))$ are 
some functions. 
\par
Let a non-associative algebra of the quantized field has 
following property
\begin{align}
  a(x) \Bigl(b(x) c(y)\Bigl) & =
  \Bigl(a(x) b(x) \Bigl) c(y)
  \label{sec2-3}\\
  \intertext{and}
  \Bigl(a(y) b(x) \Bigl) c(x) & =
  a(y) \Bigl(b(x) c(x)\Bigl)
  \label{sec2-3a}
\end{align}
here $a, b$ and $c$ are some operators depending on
the field operators $\varphi (x)$ (or $\varphi^* (x)$) 
and $\varphi (y)$ (or $\varphi^* (y)$) 
respectively. In some way this property is similar 
to the alternative property \eqref{sec1-4}. 
If $a(x) = \varphi (x)$, 
$b(x) = \varphi^* (x)$ and $c(y) = \varphi (y)$ then
\begin{equation}
  \varphi(x) \Bigl(\varphi ^*(x) \varphi(y)\Bigl) = 
  \Bigl(\varphi (x) \varphi^* (x) \Bigl) \varphi (y) .
  \label{sec2-4}
\end{equation}
Analogously 
\begin{equation}
  \Bigl(\varphi(y) \varphi ^*(x) \Bigl)\varphi(x) = 
  \varphi (y) \Bigl(\varphi^* (x) \varphi (x)\Bigl) .
  \label{sec2-4a}
\end{equation}
If $y=x$ then
\begin{equation}
  \Bigl(a(x) b(x) \Bigl) c(x) =
  a(x) \Bigl(b(x) c(x)\Bigl) =
  a(x) b(x) c(x) ,
  \label{sec2-5}
\end{equation}
\textit{i.e.} the property \eqref{sec2-5} leads to the
associativity of quantized fields in any point $x$.
\par
To determine the associators we should have
some dynamic law. In 50's Heisenberg has offered
a method of quantization of strongly nonlinear fields.
He proposed that the dynamic equation for the field operators
is the field equation for the classical field where 
the classical field is replaced by the field operator :
$\varphi(x) \rightarrow \hat{\varphi}(x)$. In this paper we 
suppose this approach can give us information to 
determine all associators. 
\par
Let us consider a simplest case of a nonlinear
spinor field with the following Hamiltonian 
(this is the Hamiltonian of the electrons system describing the
properties of a metal in the superconductivity state \cite{abr}) 
\begin{equation}
    \hat{H} = \int
    \left [
        -\left (
        \hat\psi^+_\alpha \frac{\nabla^2}{2m}\hat\psi_\alpha
         \right ) + \frac{\lambda}{2}
         \left (
         \hat\psi^+_\beta \left (
                          \hat\psi^+_\alpha \hat\psi_\alpha
                          \right )
         \hat\psi_\beta
         \right )
    \right ] d V ,
\label{sec2-6}
\end{equation}
where $\hat\psi_\alpha$ is the operator of spinor field describing
electrons; $m$ is the electron mass; $\lambda$ is some constant
and $\alpha , \beta$ are the spinor indices. According to Heisenberg
the operators $\hat\psi$ and $\hat\psi^+$ obey the following
operator equations
\begin{eqnarray}
    \left (
    i\frac{\partial}{\partial t} + \frac{\nabla ^2}{2m}
    \right )
    \hat\psi_\alpha (x) - \lambda
    \left (
    \hat\psi^+_\beta (x) \hat\psi_\beta (x)
    \right )\hat\psi_\alpha (x) & = & 0 ,
    \label{sec2-7}\\
    \left (
    i\frac{\partial}{\partial t} - \frac{\nabla ^2}{2m}
    \right )
    \hat\psi^+_\alpha (x) + \lambda
    \hat\psi^+_\alpha (x)
    \left (
    \hat\psi^+_\beta (x)
    \hat\psi_\beta (x)
    \right )& = & 0 .
\label{sec2-8}
\end{eqnarray}
As well as in Heisenberg's method for nonlinear spinor field
we have an equation for the 2-point Green's function
$G_{\alpha\beta}(x,x') = -i \langle T(\hat\psi_\alpha (x)
\hat\psi^+_\beta (x'))\rangle$
\begin{equation}
    \left (
    i\frac{\partial}{\partial t} + \frac{\nabla ^2}{2m}
    \right )G_{\alpha\beta}(x,x') +
    i\lambda  \langle T
    \left (
    \hat\psi^+_\gamma (x)\hat\psi_\gamma (x)
    \hat\psi_\alpha (x) \hat\psi^+_\beta (x')
    \right )\rangle = \delta (x - x') .
\label{sec2-9}
\end{equation}
Further we have to write an equation for term
$\langle T (\hat\psi^+_\gamma (x)\hat\psi_\gamma (x)
\hat\psi_\alpha (x) \hat\psi^+_\beta (x') )\rangle$ and so on. 
After this we will have an infinite equations set for all Green's
functions. In the textbook \cite{abr} it was made the following
approximation: the operator 
$\hat\psi_\alpha(x_1) \hat\psi_\beta(x_2)
\hat\psi^+_\gamma(x_3) \hat\psi^+_\delta(x_4)$ 
contains terms corresponding to the
annihilation and creation of bound pairs (Cooper pairs) that 
allows us to write Eq. \eqref{sec1-7}. 
We should note that expression \eqref{sec1-7} is 
approximate one and following to Heisenberg's idea 
we can cut off the above-mentioned 
infinite equations set for Green's functions. 
The last term of expression \eqref{sec1-7} plays the key role here. 
This expression gives us possibility to split the
4-point Green's function by the nonperturbative way
without using Feynman diagram techniques.
\par
After some simplifications (for details, see Ref. \cite{abr})
we can obtain two equations
\begin{eqnarray}
    \left (
    i\frac{\partial}{\partial t} + \frac{\nabla ^2}{2m}
    \right )G(x - x') - i\lambda F(0+)
    F^+(x - x') = \delta(x - x') ,
\label{sec2-11}\\
    \left (
    i\frac{\partial}{\partial t} - \frac{\nabla ^2}{2m} - 2\mu
    \right )F^+(x - x') + i\lambda F^+(0+)
    G(x - x') = 0 
\label{sec2-12}
\end{eqnarray}
where
\begin{eqnarray}
    e^{-2i\mu t} F_{\alpha\beta}(x - x') & = &
    \left \langle N \left\vert
    T \left (
    \hat\psi_\alpha (x) \hat\psi_\beta (x')
    \right ) \right\vert N + 2
    \right \rangle ,
\label{sec2-13} \\
    e^{2i\mu t} F^+_{\alpha\beta} (x - x') & = &
    \left \langle N +2 \left\vert
    T \left (
    \hat\psi^+_\alpha (x) \hat\psi^+_\beta (x')
    \right ) \right\vert N
    \right \rangle ,
\label{sec2-16}\\
    F_{\alpha\beta}(0+) & = &
    e^{2i\mu t}
    \left \langle
    N \left\vert \hat\psi_\alpha(x) \hat\psi_\beta(x) \right\vert N +2
    \right \rangle ,
\label{sec2-17}\\
    F^+_{\alpha\beta}(0+) & = &
    e^{-2i\mu t}
    \left \langle
    N+2\left\vert \psi^+_\alpha(x) \psi^+_\beta(x) \right\vert N
    \right \rangle ,
\label{sec2-18}\\
    F_{\alpha\beta}(x - x') & = & a_{\alpha\beta} F(x-x') , 
\label{sec2-18a}\\
    G_{\alpha\beta}(x - x') & = & b_{\alpha\beta} G(x-x') 
\label{sec2-18b}
\end{eqnarray}
here $a_{\alpha\beta}$ and $b_{\alpha\beta}$ are some constant 
matrixes. Now we can compare this method with 
the non-associative quantization
procedure described above. Let us suppose that the operators
$\hat\psi(x)$ form the non-associative algebra and Eq.\eqref{sec2-9} 
has following form
\begin{equation}
\begin{split}    
    \left (
    i\frac{\partial}{\partial t} + \frac{\nabla ^2}{2m}
    \right )G_{\alpha\beta}(x,x') +
    i\lambda  \left \langle 0 \left |
    \hat\psi^+_\gamma (x)\biggl(\hat\psi_\gamma (x) 
    \Bigl(\hat\psi_\alpha (x) \hat\psi^+_\beta (x')
    \Bigl)\biggl) \right | 0 
    \right \rangle \\
    = \delta (x - x') 
\label{sec2-19}
\end{split}
\end{equation}
here $|0\rangle$ is the vacuum state and 
we consider the case $x^0 > {x'}^0$ since it is not very clear 
what is the time ordering operator for the non-associative algebra. 
It can be shown (see, Appendix \ref{app2}) that 
\begin{equation}
\begin{split}
  \hat\psi^+_{\gamma}(x)\biggl(\hat\psi_\gamma(x)\Bigl(
  \hat\psi_\alpha(x) \hat\psi^+_\beta(y) \Bigl) \biggl) = 
  \Bigl(
  \hat\psi_\gamma(x) \hat\psi_\alpha(x)
  \Bigl)
  \Bigl(
  \hat\psi^+_\gamma(x) \hat\psi^+_\beta(y)
  \Bigl) \\
  - \hat\psi_\gamma(x) \hat\psi^+_\beta(y) \Delta_{\gamma\alpha}(x,x) + 
  \hat\psi_\alpha(x) \hat\psi^+_\beta(y) \Delta_{\gamma\gamma}(x,x) \\ 
  = \Bigl(
  \hat\psi^+_\gamma(x) \hat\psi^+_\beta(y)
  \Bigl)
  \Bigl(
  \hat\psi_\gamma(x) \hat\psi_\alpha(x)
  \Bigl) \\ 
  - \hat\psi^+_\gamma(x) \hat\psi_\alpha(x) \Delta_{\gamma\beta}(x,y) + 
  \hat\psi^+_\gamma(x) \hat\psi_\gamma(x) \Delta_{\alpha\beta}(x,y) \\ 
  + \hat\psi^+_\gamma(x) Ass\left(\hat\psi_\gamma(x), \hat\psi^+_\beta(y), 
  \hat\psi_\alpha(x)\right) 
  - Ass\left(\hat\psi^+_\gamma(x), \hat\psi^+_\beta(y), 
  \hat\psi_\gamma(x)\hat\psi_\alpha(x)\right) .
\label{sec2-20}
\end{split}
\end{equation}
Taking into account that $\hat\psi_\alpha(x) |0\rangle = 0$ 
we have the following equation for expression  
of the Green's function by the associator 
\begin{equation}
\begin{split}
  \left\langle 0 \left|
  \hat\psi^+_{\gamma}(x)\biggl(\hat\psi_\gamma(x)\Bigl(
  \hat\psi_\alpha(x) \hat\psi^+_\beta(y) \Bigl) \biggl)
  \right| 0 \right\rangle \\ 
  = \left\langle 0 \left|
  \Bigl(
  \hat\psi_\gamma(x) \hat\psi_\alpha(x)
  \Bigl)
  \Bigl(
  \hat\psi^+_\gamma(x) \hat\psi^+_\beta(y)
  \Bigl)
  \right| 0 \right\rangle + \text{(\textit{combination 
  of commutators})} \\
  = \left\langle 0 \left|
  \hat\psi^+_\gamma(x) 
  \right| 0 \right\rangle  
  Ass\left(\hat\psi_\gamma(x), \hat\psi^+_\beta(y), 
  \hat\psi_\alpha(x)\right) \\
  - Ass\left(\hat\psi^+_\gamma(x), \hat\psi^+_\beta(y), 
  \hat\psi_\gamma(x)\hat\psi_\alpha(x)\right) +  
  \text{(\textit{combination of commutators})} . 
\label{sec2-21}
\end{split}
\end{equation}
If all fermions are combined into pairs then we have 
$\left\langle 0 \left| \hat\psi^+_\gamma(x) \right| 0 \right\rangle =0 $. 
In Ref.~\cite{abr} it is shown that we can omit the 
\textit{combination of commutators} since it 
leads to an additive correction to the chemical potential 
in the equations for the functions $G, F, F^+$. 
Like to Eq. \eqref{sec1-7} (if $x_1 = x_2 = x_3 = x$ and 
$x_4 = x'$) we can assume that (with an accuracy of 
\textit{combination of commutators}) 
\begin{equation}
\begin{split}
    \left\langle 0 \left|
    \hat\psi^+_{\gamma}(x)\biggl(\hat\psi_\gamma(x)\Bigl(
    \hat\psi_\alpha(x) \hat\psi^+_\beta(y) \Bigl) \biggl)
    \right| 0 \right\rangle \\
    = \left \langle 0 \left |
    \Bigl(\hat\psi_\gamma (x)\hat\psi_\alpha (x) \Bigl)
    \Bigl(\hat\psi^+_\gamma (x) \hat\psi^+_\beta (x')
    \Bigl )\right| 0 \right \rangle \\
    = -Ass\left(\hat\psi^+_\gamma(x), \hat\psi^+_\beta(y), 
    \hat\psi_\gamma(x)\hat\psi_\alpha(x)\right) \\
    \approx 
    \left \langle N \left\vert
    \hat\psi_\gamma (x) \hat\psi_\alpha (x) 
    \right\vert N + 2
    \right \rangle
    \left \langle N +2 \left\vert
    \hat\psi^+_\gamma (x) \hat\psi^+_\beta (x') 
    \right\vert N
    \right \rangle .
\label{sec2-22}
\end{split}
\end{equation}
This equation allows us to derive Eqs.\eqref{sec2-11} 
\eqref{sec2-12} and others like to Green's function method 
in the superconductivity theory.  

\section{RENORMALIZATION PROCEDURE AND HEISENBERG'S 
QUANTIZATION \\ METHOD}

After 40 years of the evolution of quantum field theory we know 
how we should quantize the fields with a small coupling constant : 
we draw the Feynman diagrams and then summarize them. On this way 
we have a big problem connected with singularities of loops. 
These singularities can be eliminated by a renormalization 
procedure. Let us compare this situation with the Heisenberg's 
quantization method. The most important achievement 
of the Heisenberg's method is the possibility to write the 
operator field equations for \textit{the interacting fields}. 
And after the Heisenberg's quantization procedure we have 
infinite equations system for the n-point Green's functions 
of \textit{the interacting fields !} This is the most important 
difference of the Heisenberg's method from the perturbative diagram 
techniques which allows us to work with the propagators and vertices 
of \textit{the noninteracting fields.} For example, if we apply the 
Heisenberg's method to Green's functions in the QED then these 
Green's functions are a result of the diagram summation and 
the renormalization procedure. Probably not without reason Feynman 
said that renormalization is a method for sweeping the infinities 
of a quantum field theory under the rug. 
\par 
It allows us to say that the non-associative algebra of quantized 
fields describes 
the propagators and n-points Green's functions $(n \geq 3)$ of the 
interacting fields. The alternative condition \eqref{sec2-3} 
\eqref{sec2-3a} can be interpreted as a consequence of the field 
interaction : if $Ass(a(x), b(x), c(y) = 0)$ and 
$Ass(a(y), b(x), c(x) = 0)$ then these fields are noninteracting. 
\par 
The largest problem in the Heisenberg's quantization procedure 
is the mathematical difficulties : how to cut off the infinite 
equations system for the Green's functions ? But it is a price which 
we should pay to avoid the renormalization procedure. Nevertheless 
there are examples of such calculations : (a) the Heisenberg's calculations 
for a nonlinear spinor field \cite{heis1} \cite{heis2}; 
(b) in the Ref. \cite{dzh3} it is shown 
that the Green's function method and the Ginzburg - Landau equation in the 
superconductivity theory are the application of the Heisenberg's method. 

\section{CONCLUSIONS}

Thus, the main goal of this paper is to show that the quantization
rules of the nonlinear fields can be changed and the Heisenberg's idea 
about the quantization of strongly interacting fields leads 
to the change of the canonical operator quantization rules. 
It is shown 
that the non-associative algebra can be used to describe the 
quantized fields with a strong nonlinearity and the associators 
can be connected with the n-point Green's functions. 
In other words 
the non-associative algebra can be considered as a candidate to a role 
of an algebra of strongly interacting fields. The alternative
version of the non-associative algebra has very similar properties with pairing
in the superconductivity theory (Cooper pair) and probably in the QCD
(quark - antiquark attached at the ends of a hypothesized flux
tube filled with the color electric field).
\par 
In Ref.~\cite{dzh3} an assumption is made that Cooper electrons in a 
High-T$_c$ superconductor is connected with each another 
much stronger than in an 
ordinary superconductor in the consequence of a phonon-phonon 
interaction, just as the gluon-gluon interaction in the QCD leads 
to the confinement. It is possible that Heisenberg's quantization 
method and the non-associative algebra give us a description 
of field operators in the QCD and High-T$_c$ superconductivity. 
\par 
In the context of this approach many interesting questions arise. 
The most interesting question is the relation between proposed approach 
and weakly coupled theories (QED, $\lambda\varphi^4$ theories and so on). 
From the physical point of view the difference between these approaches 
is connected with the possibile expansion of 
associator into a series in terms of coupling constant. If such possibility 
be realized then the associator can be written as a series of Feynman diagrams. 
Another question is connected with the ordinary questions from any quantum 
field theory : is given theory unitary, causal, local and Lorentz invariant ? 
Evidently these properties of quantum field theory depend on the Lagrangian 
but not on the quantization rules. An investigation of this subject is much 
more complicated task in the case of non-associative algebra of field operators. 

\appendix
\section{THE FIRST EXAMPLE OF \\ NON-ASSOCIATIVE CALCULATIONS}
\label{app1}

Eq. \eqref{sec1-3} obtaining 
\begin{equation}
\begin{split}
  & \Bigl(ab\Bigl) \Bigl(cd\Bigl) =  
  a \Bigl( b \Bigl( cd \Bigl) \Bigl) + 
  Ass\Bigl( a,b,cd \Bigl) \\ 
  & = a \Bigl( \Bigl( bc \Bigl) d \Bigl) - 
  a Ass\Bigl( b,c,d \Bigl) + 
  Ass\Bigl( a,b,cd \Bigl) \\ 
  & = a \Bigl( \Bigl( cb \Bigl) d \Bigl) - 
  a Ass\Bigl( b,c,d \Bigl) + 
  Ass\Bigl( a,b,cd \Bigl) + 
  a \Bigl( \Delta_{bc} d \Bigl) \\ 
  & = a \Bigl( c \Bigl( b d \Bigl) \Bigl) - 
  a Ass\Bigl( b,c,d \Bigl) + 
  Ass\Bigl( a,b,cd \Bigl) + 
  a \Bigl( \Delta_{bc} d \Bigl) + 
  a Ass\Bigl( c,b,d \Bigl) \\
  & = a \Bigl( c \Bigl( db \Bigl) \Bigl) - 
  a Ass\Bigl( b,c,d \Bigl) + 
  Ass\Bigl( a,b,cd \Bigl) + 
  a \Bigl( \Delta_{bc} d \Bigl) + 
  a Ass\Bigl( c,b,d \Bigl) \\
  & + a \Bigl( c \Delta_{bd} \Bigl) \\ 
  & \Bigl( a c \Bigl) \Bigl( db \Bigl) - 
  a Ass\Bigl( b,c,d \Bigl) + 
  Ass\Bigl( a,b,cd \Bigl) + 
  a \Bigl( \Delta_{bc} d \Bigl) + 
  a Ass\Bigl( c,b,d \Bigl) \\
  & + a \Bigl( c \Delta_{bd} \Bigl) - Ass\Bigl( a,c,db \Bigl) \\ 
  & = \Bigl( ca \Bigl) \Bigl( db \Bigl) - 
  a Ass\Bigl( b,c,d \Bigl) + 
  Ass\Bigl( a,b,cd \Bigl) + 
  a \Bigl( \Delta_{bc} d \Bigl) + 
  a Ass\Bigl( c,b,d \Bigl) \\
  & + a \Bigl( c \Delta_{bd} \Bigl) - Ass\Bigl( a,c,db \Bigl) + 
  \Delta_{ac} \Bigl( db \Bigl) \\ 
  & = \Bigl( \Bigl( ca \Bigl) d \Bigl) b - 
  a Ass\Bigl( b,c,d \Bigl) + 
  Ass\Bigl( a,b,cd \Bigl) + 
  a \Bigl( \Delta_{bc} d \Bigl) + 
  a Ass\Bigl( c,b,d \Bigl) \\
  & + a \Bigl( c \Delta_{bd} \Bigl) - Ass\Bigl( a,c,db \Bigl) + 
  \Delta_{ac} \Bigl( db \Bigl) - Ass\Bigl( ca,d,b \Bigl) \\ 
  & = \Bigl( c \Bigl( ad \Bigl) \Bigl) b - 
  a Ass\Bigl( b,c,d \Bigl) + 
  Ass\Bigl( a,b,cd \Bigl) + 
  a \Bigl( \Delta_{bc} d \Bigl) + 
  a Ass\Bigl( c,b,d \Bigl) \\
  & + a \Bigl( c \Delta_{bd} \Bigl) - Ass\Bigl( a,c,db \Bigl) + 
  \Delta_{ac} \Bigl( db \Bigl) - Ass\Bigl( ca,d,b \Bigl) + 
  Ass\Bigl( c,a,d \Bigl) b \\ 
  & = \Bigl( c \Bigl( da \Bigl) \Bigl) b - 
  a Ass\Bigl( b,c,d \Bigl) + 
  Ass\Bigl( a,b,cd \Bigl) + 
  a \Bigl( \Delta_{bc} d \Bigl) + 
  a Ass\Bigl( c,b,d \Bigl) \\
  & + a \Bigl( c \Delta_{bd} \Bigl) - Ass\Bigl( a,c,db \Bigl) + 
  \Delta_{ac} \Bigl( db \Bigl) - Ass\Bigl( ca,d,b \Bigl) + 
  Ass\Bigl( c,a,d \Bigl) b \\ 
  & + \Bigl( c\Delta_{ad} \Bigl) b \\ 
  & = \Bigl( \Bigl( cd \Bigl) a \Bigl) b - 
  a Ass\Bigl( b,c,d \Bigl) + 
  Ass\Bigl( a,b,cd \Bigl) + 
  a \Bigl( \Delta_{bc} d \Bigl) + 
  a Ass\Bigl( c,b,d \Bigl) \\
  & + a \Bigl( c \Delta_{bd} \Bigl) - Ass\Bigl( a,c,db \Bigl) + 
  \Delta_{ac} \Bigl( db \Bigl) - Ass\Bigl( ca,d,b \Bigl) + 
  Ass\Bigl( c,a,d \Bigl) b \\ 
  & + \Bigl( c\Delta_{ad} \Bigl) b - Ass\Bigl( c,d,a \Bigl) b \\
  & = \Bigl( cd \Bigl) \Bigl( a b \Bigl) - 
  a Ass\Bigl( b,c,d \Bigl) + 
  Ass\Bigl( a,b,cd \Bigl) + 
  a \Bigl( \Delta_{bc} d \Bigl) + 
  a Ass\Bigl( c,b,d \Bigl) \\
  & + a \Bigl( c \Delta_{bd} \Bigl) - Ass\Bigl( a,c,db \Bigl) + 
  \Delta_{ac} \Bigl( db \Bigl) - Ass\Bigl( ca,d,b \Bigl) + 
  Ass\Bigl( c,a,d \Bigl) b \\ 
  & + \Bigl( c\Delta_{ad} \Bigl) b - Ass\Bigl( c,d,a \Bigl) b + 
  Ass\Bigl( cd,a,b \Bigl) . 
\label{app2-1}
\end{split}
\end{equation}
In these calculations we used following expressions 
\begin{eqnarray}
  \Bigl(ab\Bigl) \Bigl(cd\Bigl) & = & 
  a \Bigl( b \Bigl( cd \Bigl) \Bigl) + 
  Ass\Bigl( a,b,cd \Bigl) , 
\label{app2-2}\\
  a \Bigl( b \Bigl( cd \Bigl) \Bigl) & = & 
  a \Bigl( \Bigl( bc \Bigl) d \Bigl) - 
  a Ass\Bigl( b,c,d \Bigl) , 
\label{app2-3}\\
  a \Bigl( \Bigl( bc \Bigl) d \Bigl) & = & 
  a \Bigl( \Bigl( cb \Bigl) d \Bigl) + 
  a \Bigl( \Delta_{bc} d \Bigl)
\label{app2-4}  
\end{eqnarray}
and similar expressions with permutations of $a,b,c,d$ operators. 

\section{THE SECOND EXAMPLE \\ OF NON-ASSOCIATIVE CALCULATIONS}
\label{app2}

Eq.\eqref{sec2-20} obtaining. Taking into 
account properties \eqref{sec2-1}, \eqref{sec2-2}, 
\eqref{sec2-3} and \eqref{sec2-3a} we have 
\begin{equation}
\begin{split}
  \psi^+_{\gamma}(x)\biggl(\psi_\gamma(x)\Bigl(
  \psi_\alpha(x) \psi^+_\beta(y) \Bigl) \biggl) = 
  \Bigl(
  \psi^+_\gamma(x) \psi_\gamma(x)
  \Bigl)
  \Bigl(
  \psi_\alpha(x) \psi^+_\beta(y) 
  \Bigl) \\
  = -\Bigl(
  \psi_\gamma(x) \psi^+_\gamma(x)
  \Bigl)
  \Bigl(
  \psi_\alpha(x) \psi^+_\beta(y) 
  \Bigl) + 
  \psi_\alpha(x) \psi^+_\beta(y) 
  \Delta_{\gamma\gamma}(x,x) \\
  = -\psi_\gamma(x) 
  \biggl(
  \psi^+_\gamma(x)
  \Bigl(
  \psi_\alpha(x) \psi^+_\beta(y) 
  \Bigl)\biggl) + 
  \psi_\alpha(x) \psi^+_\beta(y) 
  \Delta_{\gamma\gamma}(x,x) \\
  = -\psi_\gamma(x) 
  \Bigl(
  \psi^+_\gamma(x)
  \psi_\alpha(x) \psi^+_\beta(y) 
  \Bigl) + 
  \psi_\alpha(x) \psi^+_\beta(y) 
  \Delta_{\gamma\gamma}(x,x) \\
  = \psi_\gamma(x) 
  \Bigl(
  \psi_\alpha(x)
  \psi^+_\gamma(x) \psi^+_\beta(y) 
  \Bigl) + 
  \psi_\alpha(x) \psi^+_\beta(y) 
  \Delta_{\gamma\gamma}(x,x) \\
  + \psi_\alpha(x)\psi^+_\beta(y) \Delta_{\gamma\gamma}(x,x) \\
  = \Bigl(
  \psi_\gamma(x) \psi_\alpha(x)
  \Bigl) \Bigl(
  \psi^+_\gamma(x) \psi^+_\beta(y) 
  \Bigl) + 
  \psi_\alpha(x) \psi^+_\beta(y) 
  \Delta_{\gamma\gamma}(x,x) \\
  + \psi_\alpha(x)\psi^+_\beta(y) \Delta_{\gamma\gamma}(x,x) .
\label{app1-1}
\end{split}
\end{equation}
On the other hand 
\begin{equation}
\begin{split}
  \psi^+_{\gamma}(x)\biggl(\psi_\gamma(x)\Bigl(
  \psi_\alpha(x) \psi^+_\beta(y) \Bigl) \biggl) = 
  \Bigl(
  \psi^+_\gamma(x) \psi_\gamma(x)
  \Bigl)
  \Bigl(
  \psi_\alpha(x) \psi^+_\beta(y) 
  \Bigl) \\
  = -\Bigl(
  \psi^+_\gamma(x) \psi_\gamma(x)
  \Bigl)
  \Bigl(
  \psi^+_\beta(y) \psi_\alpha(x) 
  \Bigl) + 
  \psi^+_\gamma(x) \psi_\gamma(x) 
  \Delta_{\alpha\beta}(x,y) \\
  = -\psi^+_\gamma(x) 
  \biggl(
  \psi_\gamma(x)
  \Bigl(
  \psi^+_\beta(y) \psi_\alpha(x) 
  \Bigl)\biggl) + 
  \psi^+_\gamma(x) \psi_\gamma(x) 
  \Delta_{\alpha\beta}(x,y) \\
  = -\psi^+_\gamma(x) 
  \biggl(\Bigl(
  \psi_\gamma(x)
  \psi^+_\beta(y) 
  \Bigl)
  \psi_\alpha(x) 
  \biggl) + 
  \psi^+_\gamma(x) Ass\left(\psi_\gamma(x), \psi^+_\beta(y), 
  \psi_\alpha(x)\right) \\
  + \psi^+_\gamma(x) \psi_\gamma(x) 
  \Delta_{\alpha\beta}(x,y) \\
  = \psi^+_\gamma(x) 
  \biggl(\Bigl(
  \psi^+_\beta(y)
  \psi_\gamma(x) 
  \Bigl)
  \psi_\alpha(x) 
  \biggl) + 
  \psi^+_\gamma(x) Ass\left(\psi_\gamma(x), \psi^+_\beta(y), 
  \psi_\alpha(x)\right) \\
  + \psi^+_\gamma(x) \psi_\gamma(x) 
  \Delta_{\alpha\beta}(x,y) - 
  \psi^+_\gamma(x) \psi_\alpha(x) \Delta_{\gamma\beta}(x,y) \\
  = \psi^+_\gamma(x) 
  \biggl(
  \psi^+_\beta(y)
  \Bigl(
  \psi_\gamma(x) 
  \psi_\alpha(x) 
  \Bigl)\biggl) + 
  \psi^+_\gamma(x) Ass\left(\psi_\gamma(x), \psi^+_\beta(y), 
  \psi_\alpha(x)\right) \\
  + \psi^+_\gamma(x) \psi_\gamma(x) 
  \Delta_{\alpha\beta}(x,y) - 
  \psi^+_\gamma(x) \psi_\alpha(x) \Delta_{\gamma\beta}(x,y)\\
  = \Bigl(\psi^+_\gamma(x) 
  \psi^+_\beta(y)
  \Bigl)\Bigl(
  \psi_\gamma(x) 
  \psi_\alpha(x) 
  \Bigl) + 
  \psi^+_\gamma(x) Ass\left(\psi_\gamma(x), \psi^+_\beta(y), 
  \psi_\alpha(x)\right) \\
  - Ass\left(
  \psi^+_\gamma(x), \psi^+_\beta(y), 
  \psi_\gamma(x) \psi_\alpha(x) 
  \right) 
  \psi^+_\gamma(x) \psi_\gamma(x) 
  \Delta_{\alpha\beta}(x,y) \\
  - \psi^+_\gamma(x) \psi_\alpha(x) \Delta_{\gamma\beta}(x,y).
\label{app1-2}
\end{split}
\end{equation}
Comparing these two equations we have 
\begin{equation}
\begin{split}
  \psi^+_{\gamma}(x)\biggl(\psi_\gamma(x)\Bigl(
  \psi_\alpha(x) \psi^+_\beta(y) \Bigl) \biggl) = 
  \Bigl(
  \psi_\gamma(x) \psi_\alpha(x)
  \Bigl)
  \Bigl(
  \psi^+_\gamma(x) \psi^+_\beta(y)
  \Bigl) \\
  - \psi_\gamma(x) \psi^+_\beta(y) \Delta_{\gamma\alpha}(x,x) + 
  \psi_\alpha(x) \psi^+_\beta(y) \Delta_{\gamma\gamma}(x,x) \\ 
  = \Bigl(
  \psi^+_\gamma(x) \psi^+_\beta(y)
  \Bigl)
  \Bigl(
  \psi_\gamma(x) \psi_\alpha(x)
  \Bigl) \\
  - \psi^+_\gamma(x) \psi_\alpha(x) \Delta_{\gamma\beta}(x,y) + 
  \psi^+_\gamma(x) \psi_\gamma(x) \Delta_{\alpha\beta}(x,y) \\ 
  + \psi^+_\gamma(x) Ass\left(\psi_\gamma(x), \psi^+_\beta(y), 
  \psi_\alpha(x)\right) 
  - Ass\left(\psi^+_\gamma(x), \psi^+_\beta(y), 
  \psi_\gamma(x)\psi_\alpha(x)\right)
\label{app1-3}
\end{split}
\end{equation}

\end{document}